\documentclass[conference]{IEEEtran}
\IEEEoverridecommandlockouts
\usepackage{cite}
\usepackage{amsmath,amssymb,amsfonts}
\usepackage{graphicx}
\usepackage{svg}
\usepackage{textcomp}
\usepackage{multirow}
\usepackage{booktabs}
\usepackage{epstopdf}
\usepackage{algorithm}
\usepackage[algo2e]{algorithm2e} 
\usepackage{algpseudocode}
\usepackage{url}
\usepackage{xcolor}
\usepackage{xcolor, soul}
\usepackage[utf8]{inputenc}
\usepackage[english]{babel}
\usepackage{listings}
\begin{document}
\title{LLM based Knowledge Graph Approach to Automating Medical Device Regulatory Compliance}
\author{\IEEEauthorblockN{Subhankar Chattoraj$^{\dagger}$, and Karuna Pande Joshi$^\dagger$} 
\IEEEauthorblockA{\textbf{$^\dagger$}\textit{Department of Information Systems, University of Maryland Baltimore County, Baltimore, MD 21250, USA}}}
\maketitle
\begin{abstract}
Advanced medical devices increasingly rely on AI-driven frameworks to automate compliance processes, ensuring safety and efficacy while reducing regulatory burdens. In the US, software-based medical devices, including those utilizing AI/ML models, are regulated by the FDA’s Center for Devices and Radiological Health (CDRH) under the Code of Federal Regulations (CFR) Title 21. These regulations are extensive, cross-referenced documents that require significant human effort to parse, leading to high compliance costs for manufacturers. We propose a novel, semantically rich framework that extracts regulatory knowledge from FDA documents and translates it into a machine-processable format. Our system encodes regulatory knowledge into an OWL/RDF based knowledge graph and uses the Mistral 7B Instruct model to dynamically generate SPARQL queries, perform compliance reasoning, and produce structured reports. This enables automated device classification (Class I, II, or III) and real time regulatory evaluation. Validated through real-world use cases, our framework significantly reduces manual review effort, enhances interpretability, and accelerates time-to-market. The proposed approach integrates AI reasoning and semantic technologies to achieve scalable, transparent, and automated regulatory compliance.
\end{abstract}
\vspace{0.2cm}
\begin{IEEEkeywords}
semantic web; knowledge graph; Large Language Model; medical device; FDA; code of federal regulations.
\end{IEEEkeywords}
\section{Introduction}
Rapid advancement of Artificial Intelligence (AI) and Machine Learning (ML) technologies is changing the landscape of medical devices \cite{Intro_1}, driving improvements in efficiency, precision, and real time decision making \cite{Intro_2,Intro_3}. Additionally, wearable medical devices are used by over 600 million people worldwide, offering benefits such as convenience, remote monitoring, and personalized healthcare, thereby improving health outcomes through use of AI \cite{Intro_61}. However, increasing reliance on AI and ML models within medical devices introduces new challenges in regulatory compliance \cite{Intro_5}, particularly as these technologies evolve faster than existing approval frameworks can accommodate. Ensuring safety, efficacy, and compliance with evolving regulations remains a critical challenge for manufacturers, regulators, and healthcare stakeholders.

In the United States, software based medical devices, including those that use AI / ML, are regulated by the Center for Devices and Radiological Health (CDRH) under the Food and Drug Administration (FDA) \cite{Intro_6}. The Code of Federal Regulations (CFR) Title 21 provides the foundational regulatory framework for medical devices, categorizing them into Class I, II, and III based on their risk levels \cite{Intro_7, Intro_8}. Although these regulations are essential to ensure patient safety and device reliability, they are highly complex, referenced, and voluminous, making compliance a time consuming and resource-intensive process \cite{Intro_9}. The manual effort required to interpret and implement these regulations leads to increased regulatory costs, potential delays in market entry, and challenges in keeping pace with AI-driven medical innovation. To address these regulatory challenges, automation technologies are increasingly being explored to streamline compliance processes. Regulatory automation can significantly reduce the manual burden by systematically parsing, interpreting, and mapping relevant regulatory clauses to device specific requirements \cite{Reg_Auto_1}. Rule based systems and knowledge graphs have already shown potential in managing static regulation sets, but with the increasing volume and dynamism of AI/ML related regulations, more adaptable and scalable solutions are needed \cite{Reg_Auto_2}. In this context, Large Language Models (LLMs) are emerging as transformative tools across multiple domains including legal, healthcare, and compliance due to their capabilities in understanding, synthesizing, and generating human like language \cite{LLM_1, LLM_2}. LLMs like GPT, Claude, and Gemini are being adopted to assist in drafting legal documents, analyzing complex regulations, and even generating programming code from natural language prompts. Their ability to reason across large, unstructured corpora makes them particularly well suited for navigating regulatory landscapes that are heavily text based and context dependent \cite{LLM_3}. However, Small language models (SLMs) offer a compelling balance of efficiency, cost-effectiveness, and task specificity.

In this work, we present a novel system that leverages  Large Language Models (LLMs) and Knowledge Graphs (KGs) to automate the regulatory compliance process for AI/ML enabled medical devices. To achieve this, we construct a semantically rich, machine processable compliance Knowledge Graph that encodes regulatory clauses, device attributes, and classification rules using linked data principles and RDF representations. Our system employs Mistral 7B Instruct, a state-of-the-art open weight small language model, to generate SPARQL queries from natural language prompts. These queries dynamically interact with the compliance KG to retrieve relevant clauses, identify classification pathways, and validate device requirements against regulatory standards. By combining LLM driven semantic querying with structured regulatory knowledge, our approach enables automated device classification and CFR clause identification, thereby reducing human effort, approval delays, and overall regulatory costs. The main contributions of this paper are as follows:
\begin{itemize}
    \item Development of a domain specific compliance knowledge graph grounded in CFR Title 21 and FDA classification schemas.
    \item Integration of a generative LLM (Mistral 7B Instruct) to produce SPARQL queries capable of navigating complex regulatory structures.
    \item Empirical evaluation of the system's ability to correctly identify classification and compliance requirements across varied medical device descriptions.
\end{itemize}
In this paper, we first discuss the related work in section II. Our semantically rich machine processable compliance KG methodology is presented in section III. The experimental evaluation and validation of the proposed KG are described in Section IV. The conclusion and future work in defined in section VI.
\begin{figure*}[!ht]
\centering
  \includegraphics[width=\textwidth]{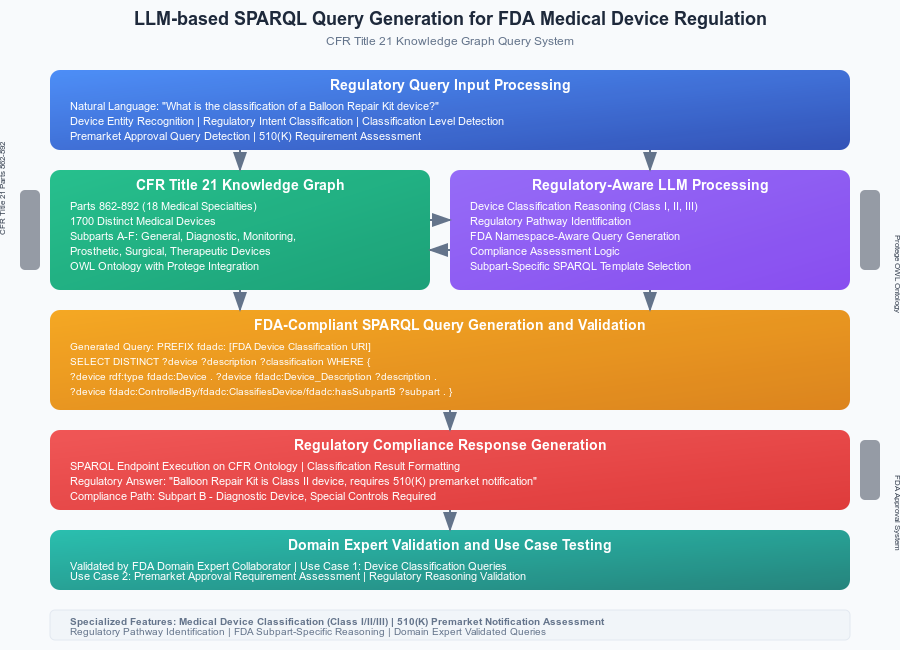}
  \caption{Overview for automating medical device regulations of the proposed semantically rich KG and LLM based architecture}
  \label{query1}
\end{figure*}
\section{Related Work}
\subsection{Regulatory Framework and Classification of Medical Devices under CFR Title 21}
The Code of Federal Regulations (CFR) represents a central body of rules issued by federal agencies under the Executive Branch of the U.S. government. It serves as an authoritative legal framework guiding regulatory compliance across domains, including healthcare and medical devices \cite{Intro_10, Intro_11}. Among its various titles, CFR Title 21, Parts 800–1050, outlines the regulatory landscape for medical devices. This includes Parts 800–861 addressing general and cross cutting requirements and Parts 862–1050 focused on device specific standards \cite{Intro_12}.

Under the Federal Food, Drug, and Cosmetic Act (FD\&C Act), the U.S. Food and Drug Administration (FDA) categorizes medical devices into three risk based classes:
\begin{itemize}
    \item Class I (low risk): Includes products like tongue depressors or stethoscopes. These devices are generally subject to minimal regulatory control.
    \item Class II (moderate risk): Covers devices such as infusion pumps or surgical drapes. These typically require compliance with special controls and the 510(k) premarket notification process \cite{Intro_13}.
    \item Class III (high risk): Encompasses critical or life supporting products like implantable pacemakers and defibrillators. These are subject to the stringent Premarket Approval (PMA) process and must provide clinical data to demonstrate safety and effectiveness \cite{Intro_14}.

\end{itemize}

For Class I and certain Class II devices, the FDA grants exemptions from 510(k) submission, provided that the devices meet specific conditions. In fact, nearly 75\% of Class I and a small subset of Class II devices fall into this exempt category, thus avoiding lengthy and expensive clinical validation \cite{Intro_15, Intro_16}. Conversely, Class III devices generally necessitate PMA due to their higher risk profile. However, in cases where the device represents only incremental innovation from a legally marketed device referred to as a predicate device manufacturers may pursue 510(k) clearance instead. Sponsors can also file a 513(g) request for classification, seeking FDA guidance on the appropriate regulatory path for novel or borderline devices \cite{Intro_17} \cite{Intro_new}.

To streamline oversight, the FDA organizes medical devices into 16 medical specialty panels, each addressing a specific therapeutic area, such as Cardiovascular, Orthopedic, or Radiology. These classifications help stakeholders navigate compliance requirements by specialty \cite{Intro_15}. In contrast to pharmaceuticals, which often require 10–12 years for market approval, medical devices typically reach the market within 3 to 7 years, depending on their risk classification and regulatory path \cite{Intro_19}. Nonetheless, the pathway to device clearance remains manual, time intensive, and non machine processable, relying heavily on expert interpretation of unstructured regulatory text \cite{Intro_20}. This imposes significant cost, resource, and time-to-market burdens on medical device sponsors especially startups and SMEs underscoring the need for automated, semantically structured compliance frameworks to accelerate approvals and reduce overhead.
\subsection{Regulatory Compliance for AI Based Medical Devices Using Semantic Web Technologies}
The integration of Artificial Intelligence (AI) and Machine Learning (ML) in healthcare is accelerating rapidly, driven by their transformative potential to enhance diagnostic accuracy, clinical decision making, and operational efficiency \cite{Intro_23, Intro_24}. Empirical studies increasingly demonstrate that AI systems can perform on par with or even surpass human experts in complex tasks such as medical image interpretation and disease prediction \cite{Intro_26}. However, this rapid adoption introduces critical regulatory, legal, and ethical challenges, particularly around safety, transparency, and accountability.

To address these issues, the U.S. Food and Drug Administration (FDA) has introduced a dedicated regulatory framework through the AI/ML Based Software as a Medical Device (SaMD) Action Plan \cite{Intro_25} \cite{Intro_new_2}. This initiative outlines strategies to ensure that AI-driven medical technologies meet rigorous standards of clinical efficacy, patient safety, and real-world reliability. For medical device developers, aligning product design and validation with such frameworks is essential to secure regulatory clearance, particularly for AI-based tools subject to 510(k) notifications or premarket approval (PMA) pathways.

The complexity of evolving healthcare regulatory requirements can be overcome by adopting Semantic Web technologies. By leveraging these technologies, we can construct a Knowledge Graph (KG) that semantically models the relationships among device classifications, regulatory codes (such as CFR Title 21), risk categories, and applicable FDA pathways. Such a KG can serve as a machine processable compliance layer, enabling automated reasoning over device eligibility, exemption status, and submission requirements. Unlike traditional document centric systems, the Semantic Web emphasizes data level interoperability. Standards such as the Resource Description Framework (RDF) \cite{Intro_22} and the Web Ontology Language (OWL) \cite{Intro_21} allow for defining rich, machine readable ontologies that represent domain specific regulatory policies. These ontologies can then support automated inference, validation, and querying using tools like SPARQL, making it feasible to encode and enforce compliance logic across distributed systems.

In this work, we aim to automate the classification of medical devices according to FDA risk tiers and CFR regulatory guidelines, using a semantically enriched KG. This not only reduces the burden of manual policy interpretation but also accelerates the device approval lifecycle ultimately lowering cost, improving time-to-market, and enhancing regulatory transparency for all stakeholders.
\subsection{KG and Semantic Technologies in Regulatory Intelligence and Post-Market Surveillance}
The growing volume and complexity of regulatory and clinical documents make efficient information retrieval a persistent challenge in regulatory science. Analysts at the FDA’s Center for Devices and Radiological Health (CDRH) often invest significant time navigating disparate sources to extract meaningful insights. To address this bottleneck, the CDRH developed the Semantic Search and Retrieval Framework (SARF), which integrates semantic technologies to enhance search accuracy and expedite access to relevant regulatory documents \cite{Intro_27}. In recent applications of Semantic Web technologies, RDF based models have been employed to detect both syntactic and logical inconsistencies in biomedical datasets, showcasing the capacity of knowledge graphs (KGs) for robust validation and reasoning \cite{Intro_28}. Similarly, SPARQL was used to examine inconsistencies in Japanese medical device regulation, demonstrating the feasibility of structured querying in regulatory compliance scenarios \cite{Intro_29}. Beyond document search and validation, Knowledge Graphs (KGs) are also transforming how we predict and monitor Adverse Drug Reactions (ADRs). In one notable study, researchers constructed a comprehensive biomedical KG consisting of entities like drugs, adverse reactions, target proteins, therapeutic indications, pathways, and genes. This KG was used to train a custom deep neural network (DNN) model, achieving promising results in ADR prediction \cite{Intro_30}. Similarly, other approaches have leveraged interlinked heterogeneous data representations in KGs to improve ADR classification by embedding semantic relationships between clinical entities \cite{Intro_31}.

Post-market surveillance plays a vital role in identifying rare or long term adverse events (AEs) that may not surface during clinical trials. Traditional surveillance systems often dependent on voluntary reporting suffer from under reporting biases. To overcome this limitation, recent research has introduced document level event extraction techniques that mine unstructured datasets to dynamically update knowledge graphs with new events and interrelations. The development of the Patient Safety Knowledge Graph (PSKG) exemplifies this approach, providing an integrated platform for synthesizing AE data from diverse surveillance repositories \cite{Intro_32}. One of the most robust sources for AE data is the FDA Adverse Event Reporting System (FAERS), which comprises over 17 million case reports, including patient diagnoses, drug reactions, medication regimens, and demographics. FAERS has also been leveraged to build disease comorbidity knowledge graphs, where edges between disease nodes represent statistically mined associations derived from rule based analysis of reported cases \cite{Intro_33}. These graphs have been evaluated using real world disease comorbidities such as psoriasis (rare), multiple sclerosis (moderate), and obesity (common) to validate the performance of automated comorbidity detection from FAERS data \cite{Intro_34}.

Together, these efforts underscore the value of KGs and semantic tools in enabling regulatory intelligence, post-market risk assessment, and automated knowledge extraction, contributing to safer and more efficient healthcare innovation pipelines.

\section{Methodology} 
In this section, we describe our framework in detail, including the techniques we used to create the KG, which captures the overall structure of CFR along with key instances. In Figure \ref{query1}, the overall architecture is illustrated. The KG is constructed using the CFR - Title 21 Part 862 - 892 comprising of the 1700 distinct devices risk category and requirement for 510(K) notification requirement or pre-market approval \cite{Intro_13}. The following section illustrates the construction and population of the CFR -Title 21 Knowledge Graph.
\subsection{CFR Title 21 Knowledge Graph Construction}
To support machine-readable regulatory reasoning, we engineered a domain specific RDF based Knowledge Graph (KG) derived from CFR Title 21, focusing on Parts 862–892, which encompass classification rules across 18 FDA-designated medical specialties. 

In designing the CFR Title 21 compliance knowledge graph, we defined several core ontology classes-Device, Classification, and Subpart each with clearly specified properties to capture FDA regulatory structure. The Device class represents individual FDA-regulated medical devices and is the central entity in the graph. Each Device instance is annotated using descriptive datatype properties such as hasDescription (a natural-language summary of its clinical function), hasProductCode (the FDA product code associated with its regulatory lineage), and hasRegulationNumber (the CFR citation under which it is regulated). Beyond descriptive metadata, devices are linked to regulatory requirements through object properties: hasClassification associates each device with an FDA risk tier (Class I, II, or III); requiresSubmission specifies the required premarket pathway (510(k), PMA, or exemption); hasSubpart maps the device to the relevant CFR subpart governing its classification or controls; and inSpecialtyPanel links the device to its FDA medical specialty panel such as Radiology or Cardiovascular.

The Classification class models the FDA’s three-tier risk hierarchy and serves as the target class for the hasClassification relation. Instances such as ClassI, ClassII, and ClassIII are individuals not subclasses and are intentionally modeled without datatype properties to reflect the categorical nature of FDA risk designations. Similarly, the Subpart class represents the structural subdivisions of CFR Title 21 (Subparts A–F). These are modeled as individuals (e.g., SubpartB, SubpartC) that act as regulatory anchor points for devices but do not themselves carry attributes beyond their identity.

It is important to clarify that Figure 2 displays instances (ABox) rather than ontology classes (TBox). Nodes such as Device\_XRay, Device\_Glucometer, ClassIII, or SubpartC are concrete individuals populated in the knowledge graph to support SPARQL-based reasoning and compliance retrieval. The conceptual classes-fdade:Device, fdade:Classification, and fdade:Subpart are not shown in the figure to maintain visual clarity \cite{Intro_new, Intro_new_2}. By visually representing instance-level relationships rather than schema-level definitions, the figure highlights how the LLM-generated SPARQL queries operate directly over real regulatory data rather than abstract class hierarchies.

\begin{figure*}[!ht]
\centering
  \includegraphics[width=\textwidth]{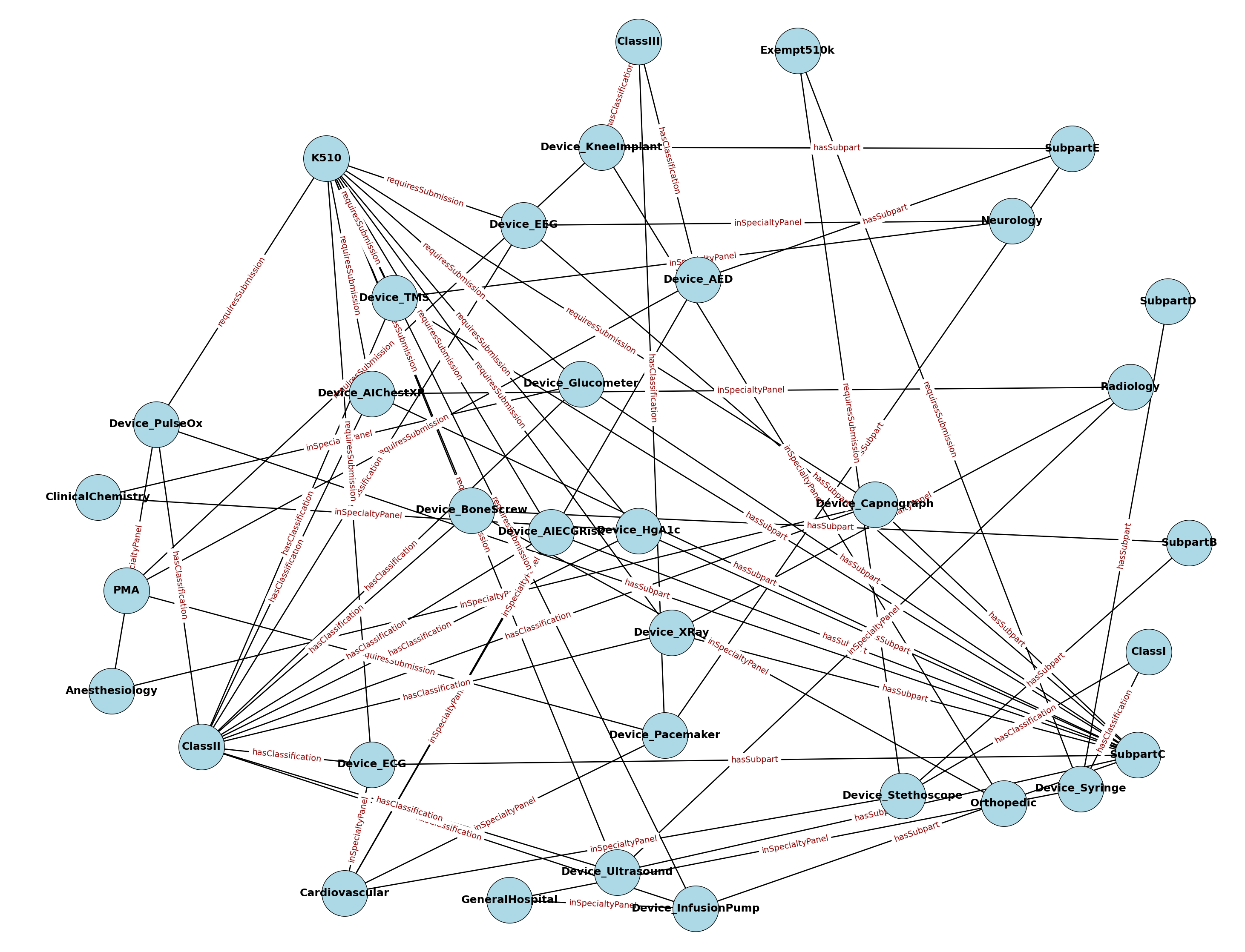}
  \caption{LLM-KG medical device Knowledge Graph top-Level Classes include Device, Subpart and Classification that are related via object properties such as hasSubpart, ControlledBy, and ClassifiesDevice}
  \label{knowledge_graph}
\end{figure*}

To enable query interoperability, the graph was developed using Protégé and serialized into RDF/XML and Turtle formats, then deployed via a SPARQL 1.1-compliant endpoint. In the Fig \ref{knowledge_graph}. LLM-KG medical device KG top-Level Classes is illustrated.
\subsection{Population of Knowledge Graph}
The graph was populated with structured instances extracted from publicly available FDA device classification databases and regulatory documents. For each device, attributes such as fdade:Device\_Description, classification level (I/II/III), regulatory subpart mapping, and applicable premarket submission types (e.g., 510(k), PMA) were extracted, normalized, and represented as RDF resources.

In addition to base population, semantic enrichment was performed by encoding classification decision logic, e.g., subpart-specific requirements for special controls or exemptions, allowing SPARQL queries to infer compliance implications. The use of typed literals, controlled vocabularies, and reified constraints enables fine-grained regulatory validation over the KG. A representative SPARQL query can identify all devices within a subpart that require 510(k) clearance and are classified as Class II, facilitating automated compliance checks.
\section{Experimental Validation}
To rigorously assess the effectiveness of our LLM-KG-based regulatory compliance system, we designed a multiphase evaluation framework that encompasses the accuracy of query generation, classification fidelity, and compliance clause retrieval. The evaluation benchmarks the system’s ability to correctly classify medical devices and retrieve applicable CFR clauses from natural language prompts.
\subsection{Dataset and Ground Truth Construction}
A validation corpus of 120 medical device descriptions was constructed to serve as ground truth for downstream evaluations. The corpus comprised 80 real-world device entries sourced from the FDA's product classification database and 40 synthetic descriptions specifically designed to simulate edge cases, such as vague or multi-functional devices. Each entry underwent manual annotation by a domain expert, who assigned FDA-recognized classification labels (Class I, II, or III), identified corresponding CFR subpart and special control requirements, and determined the appropriate premarket submission pathway, including 510(k), PMA, or exempt status. This curated dataset provided a comprehensive foundation for evaluating system performance across diverse device types and regulatory scenarios.

\subsection{Performance Metrics and Evaluation}
We define the following metrics to evaluate the end-to-end system: Query Accuracy (QA): Whether the SPARQL query generated matches the correct structure and intent. Classification Accuracy (CA): Whether the system correctly identifies the device’s regulatory class. Clause Retrieval Precision (CRP): Precision in retrieving the correct CFR subpart and compliance clauses. Compliance Resolution Accuracy (CRA): Whether the final generated answer (e.g., “Class II, 510(k), Subpart B”) matches the ground truth.

While the system achieves a strong overall performance with 89.1\% CA, further analysis of the 10.9\% misclassified cases reveals several contributing factors beyond merely “borderline or ambiguous device descriptions.” First, many errors stemmed from insufficient contextual detail in the device descriptions entries often lacked explicit information on intended use, invasiveness, or patient interaction level, which are key determinants of FDA classification. Second, semantic overlap between regulatory definitions (e.g., between certain Class I exempt and Class II 510(k) categories) led to uncertainty in the LLM’s reasoning process when explicit rule-based differentiators were absent. Third, a subset of errors arose from incomplete linkage in the underlying knowledge graph, where certain CFR clauses lacked well-defined relationships to corresponding device types, limiting the accuracy of automated inference.
\begin{table}[!ht]
\centering
\caption{System Performance Metrics}
\label{tab:performance_metrics}
\begin{tabular}{@{}ll@{}}
\midrule
\begin{tabular}[x]{@{}c@{}}\textbf{Metric}\end{tabular} & \begin{tabular}[x]{@{}c@{}}\textbf{Value (\%)} \end{tabular}\\
\midrule
Query Accuracy (QA) & 92.5 \\
\midrule
Classification Accuracy (CA) & 89.1 \\
\midrule
Clause Retrieval Precision (CRP) & 90.3 \\
\midrule
Compliance Resolution Accuracy & 87.6 \\
\midrule
\end{tabular}
\end{table}
Future iterations will address these gaps by (i) augmenting the training corpus with richer device metadata and FDA predicate examples, (ii) expanding the ontology to better capture context-dependent decision rules, and (iii) introducing a confidence-based reasoning layer to flag uncertain classifications for human review. These refinements are expected to further enhance robustness and interpretability in borderline classification scenarios.
\subsection{Baseline Comparisons}
To evaluate our system's performance, we compared it against two distinct baseline approaches representing different methodological paradigms in medical device classification. The first baseline implements a Rule-Based Classifier that utilizes hard-coded decision trees directly operationalizing the FDA's established classification framework through deterministic logic structures. This system employs keyword-based classification algorithms that scan device descriptions for specific terms corresponding to regulatory categories, while incorporating the FDA's official device taxonomy including product codes, device classes, and regulatory pathways. 

The second baseline uses an LLM-Only Zero-Shot Classification approach with Mistral 7B, operating without SPARQL integration or knowledge graph connectivity. This generative model relies entirely on pre-trained knowledge embedded within its parameters, generating classification responses through carefully crafted prompts without access to external databases or real-time regulatory information. These baselines provide comprehensive evaluation benchmarks, representing traditional expert system approaches versus modern generative AI methodologies.
\begin{table}[!ht]
\centering
\caption{Model Performance Comparison}
\label{tab:model_comparison}
\begin{tabular}{@{}lll@{}}
\midrule
\begin{tabular}[x]{@{}c@{}}\textbf{Model}\end{tabular} & \begin{tabular}[x]{@{}c@{}}\textbf{CA (\%)} \end{tabular} & \begin{tabular}[x]{@{}c@{}}\textbf{CRA (\%)} \end{tabular}\\
\midrule
Rule-Based Classifier & 72.4 & 68.3 \\
\midrule
LLM-Only Zero-Shot & 79.5 & 75.1 \\
\midrule
\textbf{Ours (LLM + KG)} & 89.1 & 87.6 \\
\midrule
\end{tabular}
\end{table}

Our hybrid LLM-KG system significantly outperformed both baselines, particularly in clause precision and formal compliance justification.

\subsection{Ablation Study}
To systematically evaluate the contribution of each architectural component, we conducted a comprehensive ablation study by systematically removing or modifying individual modules within our proposed system. The full system configuration, incorporating the large language model with knowledge graph integration and subpart-specific templates, achieved optimal performance with 89.1\% CA and 87.6\% CRA. When subpart-specific SPARQL templates were removed, performance declined substantially to 81.6\% CA and 78.9\% CRA, demonstrating the critical importance of structured query generation for precise regulatory information retrieval. The most significant performance degradation occurred when operating without knowledge graph integration (LLM-Only configuration), dropping to 79.5\% CA and 75.1\% CRA, highlighting the essential role of external knowledge augmentation in medical device classification tasks. Removing the validation layer resulted in moderate performance reduction to 83.2\% CA and 80.4\% CRA, indicating its importance for ensuring classification consistency and regulatory compliance. These ablation results conclusively demonstrate that subpart-specific SPARQL templates and the regulatory ontology constitute critical architectural components for achieving accurate clause identification and maintaining classification fidelity. The systematic performance degradation observed when removing either component validates our hybrid architectural design, confirming that the integration of structured knowledge representation with large language model capabilities is essential for robust medical device classification performance.
\begin{table}[!ht]
\centering
\caption{Ablation Study Results}
\label{tab:ablation_study}
\begin{tabular}{@{}lll@{}}
\midrule
\begin{tabular}[x]{@{}c@{}}\textbf{Configuration}\end{tabular} & \begin{tabular}[x]{@{}c@{}}\textbf{CA (\%)} \end{tabular} & \begin{tabular}[x]{@{}c@{}}\textbf{CRA (\%)} \end{tabular}\\
Without Subpart Templates & 81.6 & 78.9 \\
\midrule
Without KG (LLM-Only) & 79.5 & 75.1 \\
\midrule
Without Validation Layer & 83.2 & 80.4 \\
\midrule
LLM + KG + Templates & 89.1 & 87.6 \\
\midrule
\end{tabular}
\end{table}

\section{Discussion}
This work presents a hybrid semantic system that leverages LLMs and KGs to automate regulatory compliance for medical devices under CFR Title 21. The proposed system demonstrates strong performance in mapping natural language device descriptions to structured SPARQL queries, enabling precise retrieval of device classification and regulatory requirements. The integration of a semantically enriched, FDA-aligned ontology ensures domain fidelity and supports subpart-specific reasoning, while the LLM serves as a generative interface for dynamic query construction. Our evaluation reveals that combining LLMs with structured ontologies significantly improves both classification accuracy and clause retrieval precision compared to rule-based systems and standalone LLMs. The inclusion of a validation layer further ensures that generated queries are not only syntactically valid but also semantically aligned with FDA logic, minimizing hallucinations and enhancing trustworthiness.

However, several challenges persist. First, the reliance on accurate entity recognition and regulatory intent classification limits performance when device descriptions are ambiguous or multi-functional. Although Mistral 7B Instruct performed competitively, larger instruction-tuned models (e.g., GPT-4 or Claude) may improve robustness in such cases. Second, the system currently supports device classification and premarket pathway identification but does not yet incorporate post-market surveillance or adverse event monitoring, which are essential components of end-to-end compliance automation. Moreover, the KG construction process, while systematic, is semi-automated and relies on the availability and quality of public FDA data. Future work should explore scalable KG enrichment via web-scale regulatory scraping and alignment with international standards (e.g., EU MDR). Additionally, incorporating explainability modules for SPARQL-to-natural language translation could further improve usability for non-technical regulators and clinical reviewers. Overall, this work establishes a foundational approach toward semantic compliance automation framework for regulatory AI in high-stakes healthcare domains.
\section{Conclusion and Future Work}
In this paper, we proposed a novel framework that integrates a domain-specific compliance Knowledge Graph with Mistral 7B Instruct to automate medical device classification and regulatory reasoning under CFR Title 21. By transforming unstructured natural language descriptions into KG and automated SPARQL queries using Mistral 7B Instruct model, the system enables machine-interpretable access to FDA regulatory logic, allowing precise classification determination and clause retrieval. Empirical results across real-world and synthetic device descriptions demonstrate high classification accuracy (89.1\%) and compliance resolution fidelity (87.6\%), validating the utility of combining symbolic KGs with generative models for regulatory automation.

Key contributions include the construction of a semantically rich, OWL-based knowledge graph aligned with FDA classification schemas, the design of subpart-specific SPARQL query templates, and the implementation of a validation layer to ensure logical and regulatory consistency in generated outputs. The proposed approach reduces manual burden in regulatory review workflows and offers a transparent, auditable path toward compliance reasoning in high-stakes healthcare domains. Future work will focus on several directions. First, we aim to extend the knowledge graph to cover additional CFR parts related to post-market surveillance, adverse event reporting, and AI/ML software as a medical device (SaMD) guidelines. Second, we plan to incorporate multi-lingual support and natural language explanations of SPARQL results to improve system interpretability for regulators and clinical stakeholders. Third, we are exploring fine-tuning smaller language models on FDA-specific regulatory corpora to further enhance query generation performance within resource-constrained environments.

Lastly, future iterations will integrate cross-jurisdictional standards such as the EU MDR and IMDRF frameworks to support international compliance workflows. This work serves as a foundation for a broader class of semantically grounded, LLM-augmented regulatory reasoning systems, applicable across domains where structured compliance is critical.
\section{Acknowledgement}
We thank Dr. Andrea Iorga for helping in the expert validation of the design of our KG. This work was partially funded by the NSF Award Number 2436549, Collaborative Research: FDT-BioTech: Aspects of Digital Twin Studies for Neuroimages.
\bibliographystyle{IEEEtran}
\bibliography{References_2}
\end{document}